\documentclass[11pt]{article}

\usepackage{mdframed}
\usepackage{booktabs}
\usepackage{multirow}
\usepackage{geometry}
\usepackage{comment}

\usepackage{color}
\usepackage{float}

\usepackage{booktabs}
\usepackage{booktabs}
\usepackage{multirow}
\usepackage{array}
\usepackage{makecell}
\usepackage{apacite}

\newcolumntype{M}[1]{>{\centering\arraybackslash}m{#1}}
\usepackage{amsfonts}
\usepackage[hidelinks]{hyperref}
\usepackage{nameref}
\usepackage{enumerate}
\usepackage{caption}
\usepackage[onehalfspacing]{setspace}

\usepackage{multibib}
\newcites{App}{References}

\usepackage{setspace}

\usepackage{amsmath}
\usepackage{amsthm}
\usepackage{amssymb}

\usepackage{relsize}
\usepackage[labelformat=simple]{subcaption}
\usepackage[format=hang]{caption}
\usepackage[section]{placeins}
\usepackage{epstopdf}
\usepackage{bbm}
\usepackage{bm}

\usepackage{sgame}
\usepackage{multirow}
\usepackage{multicol}
\usepackage{array}

\usepackage{lscape}

\usepackage{float}
\usepackage{morefloats}
\usepackage{graphicx}
\usepackage{natbib}

\usepackage{soul}

\usepackage[dvipsnames]{xcolor}
\definecolor{green}{HTML}{009688}

\usepackage[ruled]{algorithm2e} 

\SetAlFnt{\small}
\SetAlCapFnt{\small}
\SetAlCapNameFnt{\small}
\SetAlCapHSkip{0pt}
\IncMargin{-\parindent}

\newtheorem{summary}{Summary}

\definecolor{shockingpink}{rgb}{0.99, 0.06, 0.75}

\definecolor{OliveGreen}{rgb}{0,0.6,0}
\makeatletter
\let\orgdescriptionlabel\descriptionlabel
\renewcommand*{\descriptionlabel}[1]{%
	\let\orglabel\label
	\let\label\@gobble
	\phantomsection
	\edef\@currentlabel{#1}%
	\let\label\orglabel
	\orgdescriptionlabel{#1}%
}
\makeatother

\begin{document}
	
\title{ GPT in Game Theory Experiments}
	\author{\Large Fulin Guo\footnote{Fulin Guo: Faculty of Economics, University of Cambridge. Email: fg400@cam.ac.uk.
 } 
}
	
\maketitle
	
\begin{abstract}

This paper explores the use of Generative Pre-trained Transformers (GPT) in strategic game experiments, specifically the ultimatum game and the prisoner's dilemma. I designed prompts and architectures to enable GPT to understand the game rules and to generate both its choices and the reasoning behind decisions. The key findings show that GPT exhibits behaviours similar to human responses, such as making positive offers and rejecting unfair ones in the ultimatum game, along with conditional cooperation in the prisoner's dilemma. The study explores how prompting GPT with traits of fairness concern or selfishness influences its decisions. Notably, the ``fair" GPT in the ultimatum game tends to make higher offers and reject offers more frequently compared to the ``selfish" GPT. In the prisoner's dilemma, high cooperation rates are maintained only when both GPT players are ``fair". The reasoning statements GPT produces during gameplay reveal the underlying logic of certain intriguing patterns observed in the games. Overall, this research shows the potential of GPT as a valuable tool in social science research, especially in experimental studies and social simulations.

\end{abstract}
	
	\setcounter{page}{0}\thispagestyle{empty}
	
	\newpage

\section{Introduction}
Experiments on strategic games have been extensively conducted to test the theoretical predictions of game theory and to explore factors that may influence the action patterns in games, such as social preferences, bounded rationality, learning, reputation building, and more.\footnote{See, for example, \cite{camerer2011behavioral}, \cite{colman2016game}, and \cite{plott2008handbook} for an overview of experimental games.} In this paper, I present a study of conducting strategic game experiments on large language models (LLM), specifically exploring the potential of the Generative Pre-trained Transformer (GPT) as a valuable tool for social science research.

The GPT is a state-of-the-art large language model developed by OpenAI that has significantly impacted the field of natural language processing and demonstrated remarkable ability in understanding and generating human-like language (\cite{chatgpt}, \cite{openai2023}). This paper presents a novel exploration of GPT's applicability in strategic game experiments.

In this study, I develop prompts to enable the GPT, specifically the GPT-4 model (\textit{gpt-4-1106-preview}), to understand game rules and engage in strategic games in a manner akin to human players. These prompts, analogous to the information displayed on a computer screen in human experiments, facilitate two GPTs in comprehending game instructions, participating in games, and engaging in repeated interactions with each other. I also prompt the GPT players to exhibit different features: one with fairness concern and the other with selfishness.\footnote{Hereafter, for the sake of simplicity, the GPT prompted to exhibit ``fairness concern" will be referred to as ``fair GPT", and the GPT prompted to exhibit ``selfishness" will be referred to as ``selfish GPT".} This setup enables an analysis of how these traits influence their behavioural patterns in the games. The GPT models are prompted to not only make decisions but also articulate the reasoning behind their choices. This approach serves a dual purpose. Firstly, research has shown that asking GPT models to think before making a decision can significantly enhance the quality of their outputs (\cite{kojima2022large}). Secondly, it provides a direct way of understanding the rationale underpinning their decisions.

This paper focuses on two simple and widely studied games in economics: the ultimatum game and the prisoner's dilemma. The ultimatum game involves a proposer who suggests splitting an amount of money between themselves and another player, and a responder who decides whether to accept or reject the proposal. If the responder accepts, the money is divided as proposed; if the responder rejects, neither player receives any money. It is well-known that the subgame perfect equilibrium (SPE) involves the proposer offering the other person the smallest possible amount and the responder accepting it. In this paper, I examine finitely repeated play of the ultimatum game over five rounds where they split a total of $100$ dollars in each round.

I find that in many important aspects, GPT's behaviours align with intuitive expectations and observed human subjects' behaviours.\footnote{See, e.g., \citet*{binmore1985testing}, \citet*{guth1982experimental}, \citet*{oosterbeek2004cultural}, \citet*{roth1991bargaining}, and \cite{thaler1988anomalies}.} For instance, backward induction is rejected, as proposers offer non-trivial amounts, and responders, particularly those with fairness concern, tend to reject low offers. Rejection rates are negatively correlated with offered amounts. The assigned traits influence the behaviours of both the proposers and the responders. On average, the fair GPT offers around $40\%$, in contrast to approximately $30\%$ by the selfish GPT. Additionally, the fair GPT exhibits a significantly higher rejection rate compared to the selfish GPT.

The reasoning statements shed light on one intriguing pattern observed in the game: the non-monotonicity in the rejection rate across rounds that peaks in the third round. Text analysis of these statements reveals that a significant fraction of GPT players accept unfair offers in the first two rounds because they anticipate that offers might be better in the future, and the increased rejection rate in the third round can be attributed to participants having faced two rounds of decreasing offers, leading them to believe that rejecting the low offers in the third round could be a necessary and effective strategy to negotiate for better future offers.

The paper then studies the prisoner's dilemma experiments on GPT. In the prisoner's dilemma, two players simultaneously decide whether to cooperate or defect. The unique Nash equilibrium involves both players defecting, but they will be better off if they both choose to cooperate. The prisoner's dilemma is one of the most studied games in economics, as it represents a classic social dilemma in which there is a discrepancy between social efficiency and individual incentives. In this paper, I examine the finitely repeated play of the prisoner's dilemma over five rounds. The payoff determination rule is that a mutual cooperation results in a payoff of $\$200 $ for both players, while a mutual defection generates a payoff of $\$100$ for both players. If one player cooperates and the other defects, the cooperator receives a payoff of $\$0$, and the defector receives $\$300 $.

The results show that while cooperation rate is significantly positive in all the scenarios, high cooperation rates can sustain only when both players are fair GPTs. That is, one selfish GPT can largely disrupt the cooperative pattern, as both fair and selfish GPT players tend to defect if the other player does so in the previous round. GPT's decisions are significantly influenced by their actions in preceding rounds, and this tendency is further shaped by their designated traits. Specifically, a fair GPT is inclined to cooperate if it defected but the opponent cooperated in the previous round. In contrast, a selfish GPT rarely cooperates in this same scenario, showing a tendency of exploiting the other player.\footnote{Significantly positive cooperation rates and conditional strategies are also observed in human participants in the prisoner's dilemma (see, e.g., \citet*{andreoni1993rational}, \citet*{cooper1996cooperation}, and \citet*{selten1986end}).}

The reasoning analysis provides insights into the motivations driving GPT's cooperation choices. In particular, I examined whether GPT's cooperation is motivated by reputation building and/or altruism, two significant drivers of cooperative behaviour in the prisoner's dilemma (\cite{andreoni1993rational}, \cite{cooper1996cooperation}, and \cite{kreps1982rational}). The findings suggest that in the first four rounds of the game, reputation building is a motivator in almost all instances of cooperation. In contrast, altruism appears to be a less significant factor, influencing less than $20\%$ of the cooperative decisions. Furthermore, the analysis of reasoning statements uncovers that cooperation, particularly in the final round, is at least partially driven by judgment errors. An example is the apparent misconception among some GPT players that their decision to defect could provoke a simultaneous defection from the other player in the same round. This finding points to a limitation in current GPT's capacity for strategic interaction and highlights the value of text analysis in uncovering the underlying logic for decisions.

Overall, this paper demonstrates the potential of conducting simple strategic game experiments using GPT models. I believe there are four primary advantages of employing LLMs, such as GPT, in strategic game experiments. Firstly, assessing whether LLMs can play games in a manner similar to humans is an important measure of AI capabilities. Secondly, conducting experiments on artificial agents is less costly than on human subjects (\citet{horton2023large}),  offering enhanced control, easy testing of different treatments, and a way to address certain ethical concerns associated with human experimentation. Thirdly, the reasoning outputs from GPT models provide insights into the rationale behind decisions, a topic of interest in behavioural economics that is often challenging to explore in human studies. Finally, this research lays the groundwork for large-scale social simulations (e.g., \cite{park2023generative}), as demonstrating human-like behaviour in GPT models at an individual level is an important step for their integration as virtual agents in simulating complex social dynamics. Such integration has the potential of offering insights into social dynamics and advancing social science theory. 
\\
\\
\textbf{Related literature.} There has been a growing body of literature conducting experiments on LLMs in social sciences (e.g., \citet*{aher2022using}, \citet*{akata2023playing}, \citet*{argyle2022out}, \citet*{bybee2023surveying}, \citet*{brand2023using}, \citet*{brookins2023playing}, \citet*{chen2023emergence}, \citet*{hagendorff2023machine}, and \citet*{horton2023large}).  In those studies, LLMs are used as proxies for human participants in surveys or experiments who respond to questions posed to them through prompts. 

For example, \citet*{horton2023large} conducts four experiments using LLMs as proxies for human participants in behavioural and labour economics, finding that the language models produce outcomes similar to those generated by human participants in these experiments. \citet*{brand2023using} show GPT-3's consistency with economic theory and consumer behaviour, including realistic willingness-to-pay estimates.

Focusing on game theory, \citet*{aher2022using} test GPT-3 in economic and social experiments like the ultimatum game, finding human-like responses from the \textit{text-davinci-002} model. \cite{akata2023playing} analyse the cooperation and coordination behaviour of LLMs in finitely repeated games, revealing their proficiency in self-interest-focused games and challenges in coordination tasks. \cite{brookins2023playing} study AI's preferences for fairness and cooperation by comparing GPT-3.5's decisions in the dictator game and the prisoner's dilemma with human behaviours, and they find that the GPT exhibits a stronger inclination towards fairness and cooperation as compared to humans.

The main distinctions of my paper lie in (1) the application of the latest GPT-4 model in strategic interaction environments, in contrast to most of the existing research which uses GPT-3.5 for game theory experiments, (2) the exploration of how GPT's behavioural patterns in games are influenced by specific characteristics assigned to it through prompts., and (3) the analysis of the reasoning behind GPT's decisions, achieved through the text analysis of its generated statements.  This research underscores the potential of GPT as a supplementary tool for traditional human experiments and provides insights into the realm of LLM-based social and economic simulations.

The remaining sections of the paper are organised as follows: Section \ref{sec:design} outlines the methodology. Section \ref{sec:ug} presents the results of the ultimatum game while Section \ref{sec:pd} shows the outcomes of the prisoner's dilemma. Section \ref{sec:con} is the discussion. The prompts used in the experiments are presented in \hyperref[app:prompts]{Appendix}.

\section{Methodology}\label{sec:design}

I conducted the experiments using the OpenAI Chat Completions API, specifically the \textit{gpt-4-1106-preview} chat model. This API facilitates the generation of responses from GPT based on prompts given in each round, which include game instructions, the GPT's assigned features (``fair" or ``selfish"), the history of previous rounds, and other relevant information. These prompts are analogous to the information displayed on a computer screen during human experiments. The exact prompts used in this paper can be found in Appendix \ref{app:prompts}. This section provides a high level summary of the key components.

Each prompt includes a system message and a user message.\footnote{The system message defines the general behaviour of the GPT model, and the user message provides specific requests for the GPT model to answer. For a description of the \textit{gpt-4-1106-preview} model, refer to the \textcolor{blue}{\href{https://platform.openai.com/docs/guides/chat}{OpenAI documentation}}.} In this work, the system message informs GPT that it will be participating in a multi-round game and advises it to pretend to be a human in the game with the assigned features. This system message sets the general context for the GPT and informs how it should behave. The user message contains all other information needed in each round, which includes the payoff determination rule, the information about the GPT's and its opponent's past choices and payoffs, and the format of outputs to be generated.

In each experiment, I consider two different assigned features: fairness concern and selfishness. The importance of these behavioural characteristics has been explored in the literature in both the ultimatum game (\citet*{fehr1999theory}) and the prisoner's dilemma (\cite{andreoni1993rational}). This differentiation is implemented through a slight modification in the prompts sent to the fair and selfish GPT. The fair GPT receives the following prompt: ``Please pretend that you are a human in the game with the following features when making decisions: payoff maximization, strategic thinking, fairness concern." Conversely, for the selfish GPT, the prompt is: ``Please pretend that you are a human in the game with the following features when making decisions: payoff maximization, strategic thinking, selfishness", with the phrase ``fairness concern" being replaced by ``selfishness".  The inclusion of ``payoff maximization" and ``strategic thinking" in both prompts ensures that the GPT models engage in the strategic games instead of generating game-irrelevant outputs, as these are fundamental elements in such games.

The prompts instruct the GPT to output both its reasoning and decision, where it is told that the ``reasoning" should briefly explain its reasoning before making the decision. The choices and reasoning statements generated by GPT players are then be analysed.
\\
\\
\textbf{Parameters.}
There are five rounds in each game where an agent plays against the same opponent across all the rounds. From the second round onwards, players are informed of the choices and payoffs made by both players in the past rounds. In the ultimatum game, the proposer and the responder divide a total of 100 dollars of money. In the prisoner's dilemma, cooperation generates a payoff of \$200 for both players while defection produces \$100 for both players. If one player cooperates and the other defects, the cooperator receives a payoff of \$0 and the defector receives \$300.

As two distinct characteristics are assigned to the GPT players: selfishness (selfish) and fairness concern (fair), there are four different treatment groups in the ultimatum game: a selfish proposer with a selfish responder, a selfish proposer with a fair responder, a fair proposer with a selfish responder, and a fair proposer with a fair responder. Similarly, there are three different treatment groups in the prisoner's dilemma: two selfish players, one selfish and one fair player, and two fair players. For each treatment group, $100$ simulations of the game were conducted, resulting in a total of $(4+3) \times 100 = 700$ observations for the entire study in this paper.\footnote{The experiments were conducted between November 16 and November 20, 2023.}

Temperature is a parameter used to control the degree of randomness and diversity in the generated outputs of GPT models. As in, e.g., \cite{brand2023using}, the temperature in this research is set to $1$ to allow a large variation in responses and the game dynamics. Other parameters are set to default.
\\
\\
\textbf{Analysis of the Reasoning Texts.} 
As the prompts instruct the GPT to provide both reasoning and decision in the outputs, the reasoning statements it generates become a key resource for understanding the rationale behind specific choices and intriguing patterns in the games. 

The process for analysing these reasoning statements may vary depending on the specific tasks, but it generally has the following components. The initial step involves identifying a range of relevant categories to which the reasoning statements might belong. These categories could encompass well-established concepts like reputation building and altruism, along with insights gleaned from preliminary statement reviews. Next, a large language model is employed for the automatic classification of each statement, identifying whether each element in the category set is present in each statement. To ensure the accuracy and reliability of the language model's classifications, a manual review is conducted on a randomly selected subset of the statements.

I use the GPT-4-preview-1106 model for this task. GPT-4 model has previously been employed in text or sentiment analysis and has demonstrated its relatively high capability in such tasks (e.g., \cite{de2023generative}, \cite{kheiri2023sentimentgpt}, \cite{rathje2023gpt}). In analysing the reasoning texts, GPT-4 is queried to identify if each text includes specific elements (e.g., reputation building). The model receives detailed contextual information about each reasoning statement, such as ``a given reasoning statement provided by a proposer in a multi-round Ultimatum Game". The GPT is also provided with the descriptions of each element in the category set. The prompts for the reasoning analysis can be found in the Github repository for this paper. Note that this task leverages the natural language processing capabilities of GPT-4 as given and evaluating GPT-4's capability in text or sentiment analysis is not the focus of this paper. The temperature is set to 0 to ensure mostly deterministic responses.

\section{Ultimatum Game}\label{sec:ug}

This section presents the results of the repeated ultimatum game. The ultimatum game involves a proposer who suggests splitting an amount of 100 dollars between themselves and another player, and a responder who decides whether to accept or reject the proposal. If the responder accepts, the money is divided as proposed; if the responder rejects, neither player receives any money. In this paper, the proposer plays against the same responder for five rounds. Recall that two distinct characteristics are assigned to the GPT players: selfishness (selfish) and fairness concern (fair). This leads to four different treatment scenarios: a selfish proposer with a selfish responder, a selfish proposer with a fair responder, a fair proposer with a selfish responder, and a fair proposer with a fair responder.

\begin{table}[h]
\centering
\caption{Summary statistics of the proposer's behaviour}
\begin{tabular}{lcccc}
\toprule
Variable   & \multicolumn{1}{c}{selfish-selfish} & \multicolumn{1}{c}{selfish-fair} & \multicolumn{1}{c}{fair-selfish} & \multicolumn{1}{c}{fair-fair} \\
\midrule
mean offer & \makecell{28.552*** \\ (0.335)} & \makecell{30.940*** \\ (0.259)}  & \makecell{38.826*** \\ (0.312)}  & \makecell{40.278*** \\ (0.266)}\\ \hline
change of offer after acceptance &  \makecell{-3.677*** \\ (0.158)} & \makecell{-3.054*** \\ (0.190)}  & \makecell{-1.605*** \\ (0.163)}   & \makecell{-0.815*** \\ (0.163)} \\ \hline
change of offer after rejection &  \makecell{6.240*** \\ (0.604)} & \makecell{4.932*** \\ (0.292)}  & \makecell{4.000* \\ (1.000)}   & \makecell{7.591*** \\ (0.844)}  \\ \hline
\bottomrule
\end{tabular}
\label{tab:ug_pro}
 \captionsetup{font=footnotesize, labelformat=empty}
  \caption*{Notes: This table shows the summary statistics of the proposer's behaviour in the ultimatum game. The results are shown separately for the four treatments: selfish proposer with selfish responder, selfish proposer with fair responder, fair proposer with selfish responder, and fair proposer with fair responder. The numbers in parentheses represent standard errors. ***: $p<0.01$, **: $p<0.05$, *:$p<0.1$.}
\end{table}

Table \ref{tab:ug_pro} shows the summary statistics of the proposer's behaviour in the ultimatum game. The first finding is that the outcomes reject the theory of subgame perfect equilibrium,\footnote{This is consistent with the experimental findings on human behaviour in ultimatum games (see, for example, \citet*{guth1982experimental}, \citet*{roth1991bargaining}, and \cite{thaler1988anomalies}, among others).} as the proposers offer significantly positive amounts of money to the responder in all treatments ($p<0.01$, see row 1 in Table \ref{tab:ug_pro}). Specifically, the average offer from a selfish GPT proposer is around 30, while it is about 40 for a fair GPT proposer. The offered amount is significantly different between the fair GPT and selfish GPT (Wilcoxon signed-rank test, $p<0.01$). In comparison to human players, a meta-study of 37 papers by \citet{oosterbeek2004cultural} reports that human proposers, on average, offer 40\% of the total amount, which is similar to the outcome of the fair GPT proposers.\footnote{The offered amount by human proposers can vary considerably across different studies (\cite{oosterbeek2004cultural}). For instance, \cite{henrich2000does} finds that the average offered amount is only $26 \%$ among Peruvian Amazon people.}

Another notable finding is GPT proposers' reactive behaviour to the outcomes of the previous round. ``Change of offer after acceptance" represents the difference in the offer compared to the last round if the previous offer was accepted. The results show that if the offer was accepted in the last round, there is a significant decrease in the next round's offer amount, with this reduction being more pronounced for selfish proposers than for fair ones. In contrast, following a rejected offer, proposers in all four treatment groups tend to increase their subsequent offer.

\begin{table}[h]
\centering
\caption{Summary statistics of the responder's behaviour}
\begin{tabular}{lcccc}
\toprule
Variable   & \multicolumn{1}{c}{selfish-selfish} & \multicolumn{1}{c}{selfish-fair} & \multicolumn{1}{c}{fair-selfish} & \multicolumn{1}{c}{fair-fair} \\
\midrule
overall rejection rate & \makecell{0.050*** \\ (0.010)} & \makecell{0.178*** \\ (0.017)}  & \makecell{0.006* \\ (0.003)}  & \makecell{0.044*** \\ (0.009)}\\ \hline
rejection rate after offer increase &  \makecell{0.000 \\ (0.000)} & \makecell{0.036** \\ (0.016)}  & \makecell{0.000 \\ (0.000)}   & \makecell{0.019 \\ (0.013)} \\ \hline
rejection rate after offer decrease &  \makecell{0.077*** \\ (0.015)} & \makecell{0.366*** \\ (0.032)}  & \makecell{0.018* \\ (0.010)}   & \makecell{0.164*** \\ (0.034)}  \\ \hline
\bottomrule
\end{tabular}
\label{tab:ug_res}
 \captionsetup{font=footnotesize, labelformat=empty}
  \caption*{Notes: This table shows the summary statistics of the responder's behaviour in the ultimatum game. The results are shown separately for the four treatments: selfish proposer with selfish responder, selfish proposer with fair responder, fair proposer with selfish responder, and fair proposer with fair responder. The numbers in parentheses represent standard errors. ***: $p<0.01$, **: $p<0.05$, *:$p<0.1$.}
\end{table}

Table \ref{tab:ug_res} presents the summary statistics of the responder's behaviour in the ultimatum game. As shown in the first row of the table, the rejection rate is significantly positive at $1\%$ level in all scenarios except for the FS treatment (a fair proposer and a selfish responder). Notably, the rejection rate in the SF treatment (a selfish proposer and a fair responder) is the highest among the four groups, approximately $18\%$. This is consistent with intuitive expectation, as a selfish proposer is more likely to make lower offers, while a responder with fairness concern tends to reject such offers, thus leading to a relatively high rejection rate for this combination. In comparison, human studies show an average acceptance rate of around $16\%$ (\cite{oosterbeek2004cultural}), similar to the ``selfish proposer-fair responder" scenario in my study. 

Additionally, analogous to the proposers' behaviours, the rejection rate is much higher when the offer decreases as compared to that following an increase in the offer. This difference is most pronounced for the SF treatment, where a fair responder rejects decreased offers with more than one-third frequency, while the rejection rate falls below $5\%$ after an offer increase. The impact of payoff change on the responder's behaviour aligns with intuitive expectation and has also been found in human subjects (\cite{cooper2011dynamics}).

\begin{figure}[H]
\caption{Offer and rejection rate per round}\begin{subfigure}[b]{.5\textwidth}
\centering
\includegraphics[scale=.49]{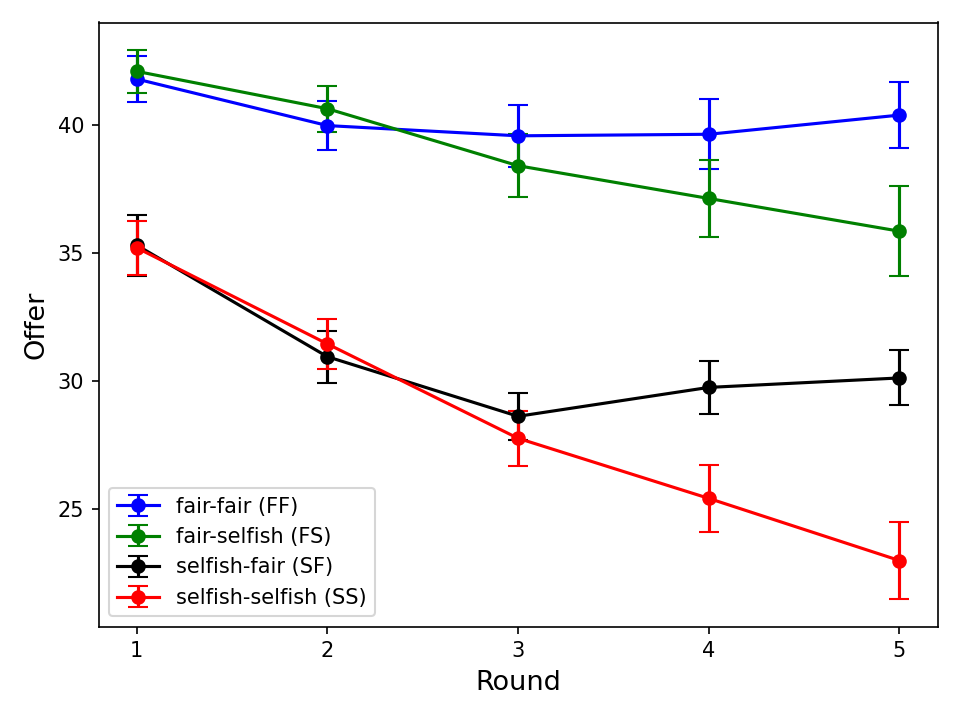}
		\vspace{-.00in}
		\caption{Mean offer vs. round}\label{fig:offer_rnd_rat_soc}
	\end{subfigure}
	\begin{subfigure}[b]{.5\textwidth}
		\centering
\includegraphics[scale=.49]{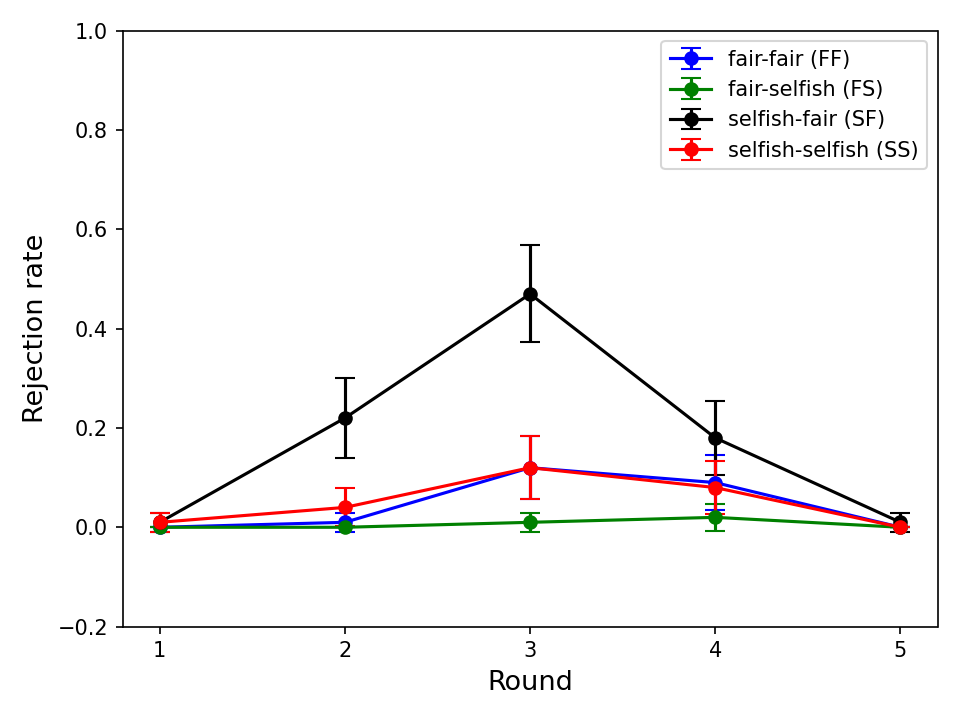}
		\vspace{-.00in}
		\caption{Rejection rate vs. round}\label{fig:reject_rnd_rat_soc}
	\end{subfigure}
	\label{fig:sta_rnd}
\end{figure}

Regarding the outcomes across rounds, Figure \ref{fig:offer_rnd_rat_soc} shows the average offer proposed across the five rounds. First, we observe that fair proposers consistently make higher offers than selfish proposers in all rounds. Furthermore, the nature of the responder influences the proposer's offer. Specifically, when paired with a selfish responder, proposers tend to make lower offers in the last two rounds, displaying a clear decreasing trend throughout the game. Conversely, when interacting with a fair responder, there is no decreasing trend in offers in the final three rounds.

These findings are also reflected in the first column of Table \ref{tab:ug_regress}, which presents an OLS regression of the offered amount against the round number and two dummy variables for whether the proposer or responder is selfish. The regression results reveal a significantly negative correlation between the offered amounts and the progression of rounds. Additionally, selfish proposers consistently offer lower amounts than fair proposers. Intriguingly, even though proposers are not informed about the character of the responder (selfish or fair) --- as this information is not included in the prompt --- the offers are significantly lower when the responder is selfish. This observation could be attributed to the fact that selfish responders, who primarily focus on maximising payoff without fairness considerations, might be more inclined to accept lower offers, which incentivises the proposers to accordingly reduce their offers.

Figure \ref{fig:reject_rnd_rat_soc} illustrates the evolution of rejection rates across rounds in the ultimatum game. Beyond the different overall rejection frequencies among the four groups, a notable finding is that the rejection rate peaks in the third round, while being almost negligible in both the first and last rounds (only 2 cases of rejection in the first round and 1 case of rejection in the last round). The near-zero rejection rate observed in round 5 may be attributed to the endgame effect. However, the relatively high rejection rate in round 3 presents an intriguing phenomenon that may not be immediately clear. This specific pattern will be explored by analysing the reasoning statements generated by GPT to understand the underlying logic behind the non-monotonic pattern of rejection frequencies throughout the game. 

The last column of Table \ref{tab:ug_regress} presents a logit regression of rejections on the offered amount, round number, and the two dummy variables for GPT traits. The results indicate that rejection rates decrease with higher offers, a finding consistent with human behaviour and intuitive expectations. Moreover, rejection rates are negatively correlated with the round number. Selfish responders are found to have significantly smaller rejection rates than fair ones, likely due to the GPT model interpreting selfishness as a focus on maximising payoffs, with less regard for equity. The selfishness of the proposer, however, does not significantly impact the rejection frequency.

\begin{table}[h]
\centering
\caption{Regression analysis of the ultimatum game}
\begin{tabular}{lcc}
\toprule
Variable   & \multicolumn{1}{c}{offered amount} & \multicolumn{1}{c}{rejection}  \\
\midrule
offered amount & \makecell{ \\ } & \makecell{-0.237*** \\ (0.021)} \\ \hline
round &  \makecell{-1.531*** \\ (0.099)} & \makecell{-0.370*** \\ (0.087)}  \\ \hline
proposer being selfish &  \makecell{-9.806*** \\ (0.279)} & \makecell{-0.245 \\ (0.280)}  \\ \hline
responder being selfish &  \makecell{-1.920*** \\ (0.279)} & \makecell{-2.539*** \\ (0.274)}  \\ \hline
constant &  \makecell{45.106*** \\ (0.382)} & \makecell{6.692*** \\ (0.888)}  \\ \hline
\bottomrule
\end{tabular}
\label{tab:ug_regress}
 \captionsetup{font=footnotesize, labelformat=empty}
  \caption*{Notes: The first column presents an OLS regression of the offered amount on the round number and two dummy variables indicating whether the proposer or responder is selfish. The second column shows a logit regression of whether rejecting on the offered amount, the round number, and the same two dummy variables. The numbers in parentheses represent standard errors. ***: $p<0.01$, **: $p<0.05$, *:$p<0.1$.}
\end{table}

\par\noindent\textbf{Reasoning Analysis}

This section focuses on analysing the reasons behind GPT's decisions by conducting text analysis on the reasoning statements it generates. Recall that the prompts direct GPT to generate both its reasoning and decision in each round when making a choice. By scrutinising these reasoning statements, the aim is to delve into the underlying reasons for any unexpected or intriguing choices or patterns that have emerged during the games.

Recall that we observed a non-monotonic pattern of rejection rates across rounds and the rejection rate is the highest in the third round among all the five rounds, especially for the combination of a selfish proposer and a fair responder. The logic driving this specific pattern may not be immediately apparent. To unravel this, I first look into the reasoning statements of GPT responders who reject offers in the third round. 

Here are some examples of such texts:

{\captionsetup{font=small, labelformat=empty}
\captionof{figure}{Reasoning statements generated by GPT responders who reject offers in round 3}
}
\begin{mdframed}[linecolor=blue, backgroundcolor=gray!20]
\small
GPT: 
\textit{``The offer is less than previous rounds, indicating a trend of decreasing offers. By rejecting, I signal unwillingness to accept such low offers, with the intention to push for a fairer split in the remaining rounds, maximizing overall payoff."}
\\
\\
GPT:
\textit{``The proposer is progressively offering a smaller share, signaling a push of boundaries. Rejecting might signal a preference for fairer offers, possibly leading to better offers in subsequent rounds. This strategic move could maximize overall payoff despite the immediate loss, and reflects a fairness concern."}
\\
\\
GPT:
\textit{``Accepting the offer would increase my total payoff, which aligns with my goal of payoff maximization. However, the offers are getting progressively worse, and by rejecting this one, I can signal to the proposer that I expect a fairer share. This strategic move might encourage a more balanced offer in future rounds, but it also risks gaining nothing this round. Given the previous pattern of decreasing offers and with two rounds left, I might encourage fairness and possibly better offers by rejecting this one."}
\end{mdframed}

We can see that two key insights are present in all the three reasoning texts above. First, there is an awareness of the proposer's increasingly unfair offers. Second, the responder considers the potential long-term benefits of rejecting the offer, anticipating that this might lead to more favourable offers in future rounds.

For the text analysis, I classify each statement based on whether it contains information about ``consistently diminishing offers" and ``potential for higher future offers." Remarkably, all 72 rejection statements in the third round indicate that rejections were made in the hope of receiving better offers in future rounds, which is a sign of strategic rejection. Furthermore, approximately 80\% of these responses mention encountering a trend of consistently decreasing or less fair offers.

To further understand the non-monotonic pattern in the rejection behaviour throughout the game, I also examine the reasoning statements in other rounds for those who accept offers no larger than 30. Here are some examples of those reasoning texts in round 1, 2, 4, and 5, respectively.

\begin{mdframed}[linecolor=blue, backgroundcolor=gray!20]
\small
A GPT responder who accepts an offer $\leq 30$ in round 1: 
\\
\textit{``Accepting the offer maximizes my immediate payoff since rejecting would yield no money. Future rounds provide an opportunity to negotiate for a better share, but securing some funds now is a pragmatic choice given my goal of payoff maximization."}
\end{mdframed}

\begin{mdframed}[linecolor=blue, backgroundcolor=gray!20]
\small
A GPT responder who accepts an offer $\leq 30$ in round 2:
\\
\textit{``The offer is less than the previous round, but still provides a positive gain. Rejecting offers no gain. With fairness concern but with strategic thinking for payoff maximization over multiple rounds, accept the offer and signal willingness to cooperate but expect fairer offers in future rounds."}
\end{mdframed}

\begin{mdframed}[linecolor=blue, backgroundcolor=gray!20]
\small
A GPT responder who accepts an offer $\leq 30$ in round 4:
\\
\textit{``Accepting the offer maximizes my immediate payoff, as rejecting would yield no money. Although the offer is not fair and I have concerns about fairness, rejecting it would not improve fairness in past or future rounds. There is only one round left, making it less likely that the proposer will adjust their behavior significantly in the final round, so the strategic benefit of rejecting to influence future offers is limited."}
\end{mdframed}

\begin{mdframed}[linecolor=blue, backgroundcolor=gray!20]
\small
A GPT responder who accepts an offer $\leq 30$ in round 5:
\\
\textit{``Even though the offer is not perfectly fair, rejecting it will result in no gain for either player, which does not align with my payoff maximization strategy. Since it is the last round, I cannot use rejection as a strategy to potentially influence future offers. Hence, I accept to ensure the highest total payoff for myself over the course of the game."}
\end{mdframed}

From the provided statements, it is evident that the GPT responders recognise the value of accepting offers to secure some payoff, as opposed to receiving nothing in the case of rejection. In the earlier rounds (1 and 2), the GPT responders exhibit an expectation of better offers or opportunity to negotiate in the future. However, in the later rounds (4 and 5), the reasoning shifts to an acknowledgment of the limited number of rounds remaining for benefiting from strategic rejection.

Building on these initial review of the reasoning texts, I categorise each statement to determine if it includes the following types of information: ``accepting provides some gains versus nothing", ``anticipation of better or fairer future offers", and ``awareness of the limited number of remaining rounds".

The results show that in the first two rounds, with 158 acceptances of offers no larger than 30, all the GPT responders note that accepting the offer, despite its low value, can result in a positive outcome, or mention that rejecting it would lead to no gain. Interestingly, approximately two-thirds of these statements express an expectation of better offers in future rounds (without the need of rejection in the current round), and less than 10\% of the GPT players think that there is a limited number of rounds remaining.

In the last two rounds, which have 322 instances of acceptance for offers smaller than or equal to 30, all responses again highlight the benefit of gaining something over nothing by accepting the offer. In contrast to the reasoning patterns in the first two rounds, only approximately 10\% of GPT responders still anticipate better future offers. Notably, around 80\% of them mention the limited number of rounds remaining, which indicates that there is little gain from strategic rejection.

These findings offer insights into the relatively high rejection rate in the third round: By this stage, participants have experienced two rounds of diminishing offers, as opposed to just one in the second round, and are likely disillusioned with the lack of improvement in proposals. With two rounds still remaining for negotiation, the third round represents a point where strategic rejection is perceived as neither too premature (first two rounds) nor too late (last two rounds).

The main findings in the ultimatum game are summarised as follows:

\begin{summary}

In the ultimatum game, GPT displays behaviours akin to those of humans. It generally makes offers that are positive yet less than half of the total amount. GPT also rejects low offers at a non-trivial rate, especially when prompted with fairness concern. Its decisions are reactive: proposers modify their offers in response to the previous decisions of responders, and the acceptance by responders is influenced by the trend of offers. The effects of assigned features to GPT are pronounced, with a fair GPT making higher offers and rejecting more frequently than a selfish GPT. The analysis of reasoning statements indicates that the relatively high rejection rate in the game's third round can be attributed to responders facing progressively diminishing offers and perceiving sufficient remaining rounds to make negotiation worthwhile.

\end{summary}

\section{Prisoner's Dilemma}\label{sec:pd}

In this section, the results of the finitely repeated prisoner's dilemma are presented. The prisoner's dilemma involves two players simultaneously choosing whether to cooperate or defect. For this study, GPT plays the prisoner's dilemma against the same opponent for five rounds. In each round, a mutual cooperation produces $\$200 $ for both players, while a mutual defection generates $\$100$ for both players. If one player cooperates and the other defects, the cooperator receives nothing, and the defector receives $\$300 $. With two distinct characteristics assigned to the GPT model --- selfishness (selfish) and fairness concern (fair) --- the experiment has three different combinations of GPT player pairings: two selfish players, one selfish and one fair player, and two fair players.

Table \ref{tab:pd_sum} provides a summary of the cooperation behaviours for each treatment. Firstly, cooperation rates are significantly greater than zero in all the treatments, which is consistent with the results of human subjects in finitely repeated prisoner's dilemma experiments (e.g., \citet*{andreoni1993rational}, \citet*{cooper1996cooperation}, and \citet*{selten1986end}). There are large effects of the assigned characteristics on cooperation frequencies: the cooperation rate when both GPTs have fairness concern is nearly $100\%$, while it drops to less than $20 \%$ when a fair GPT is paired with a selfish GPT, indicating that just one selfish GPT player is sufficient to largely disrupt the cooperative pattern. Moreover, the cooperation rate is less than $10\%$ when both GPT players are selfish.

\begin{table}[h]
\centering
\caption{Summary statistics of the prisoner's dilemma}
\begin{tabular}{lccc}
\toprule
Variable   & \multicolumn{1}{c}{selfish-selfish} & \multicolumn{1}{c}{selfish-fair}  & \multicolumn{1}{c}{fair-fair} \\
\midrule
rate of choosing C & \makecell{0.091*** \\ (0.009)} & \makecell{0.189*** \\ (0.012)} & \makecell{0.994*** \\ (0.002)} \\ \hline
rate of CC &  \makecell{0.026*** \\ (0.007)} & \makecell{0.022*** \\ (0.007)} & \makecell{0.988*** \\ (0.005)} \\ \hline
rate of CD and DC &\makecell{0.130*** \\ (0.015)} & \makecell{0.334*** \\ (0.021)} & \makecell{0.012** \\ (0.005)} \\ \hline
rate of DD  & \makecell{0.844*** \\ (0.016)} & \makecell{0.644*** \\ (0.021)} & \makecell{0.000 \\ (0.000)} \\ \hline
\bottomrule
\end{tabular}
\label{tab:pd_sum}
 \captionsetup{font=footnotesize, labelformat=empty}
  \caption*{Notes: the variables displayed are: rate of choosing C (cooperation), rate of CC (both players cooperating), rate of CD and DC (one player cooperating and the other defecting), and rate of DD (both players defecting). The results are shown separately for the three treatment groups: two selfish players, one selfish and one fair player, and two fair players. The numbers in parentheses represent standard errors. ***: $p<0.01$, **: $p<0.05$, *:$p<0.1$.}
\end{table}

\begin{figure}[H]
\captionsetup{skip=-1pt}
\caption{Cooperation rate per round}
\centering
\includegraphics[width=0.7\linewidth]{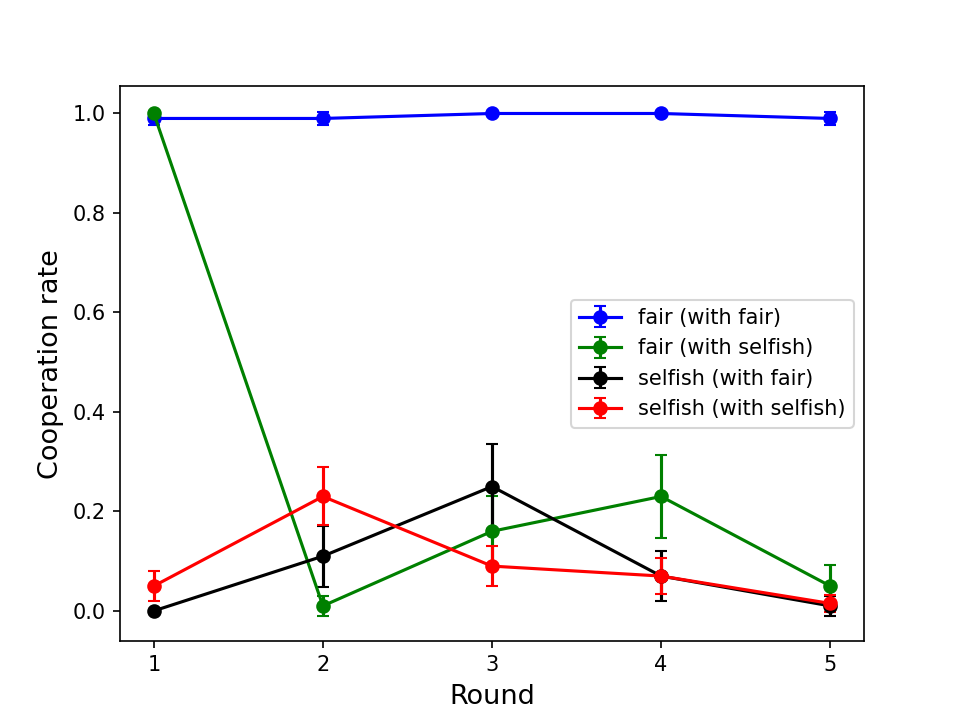}
\label{fig:coop_rnd}

\end{figure}

Figure \ref{fig:coop_rnd} shows the evolution of cooperation rates across rounds, differentiating between fair and selfish GPT players, and their respective pairings. In the first round, the fair GPT almost always cooperates and the selfish GPT mostly defects. However, in subsequent rounds, the cooperation rates remain high (near one) only if the fair player is paired with another fair player and they are small (below 25\%) in all the other three cases, including where the fair GPT player is paired with a selfish GPT player.

Given that the GPT players are not informed about their opponent's type, it is logical that the fair GPT exhibits similar behaviour in the first round regardless of the opponent's character. The significant drop in cooperation rates in subsequent rounds suggests that the fair GPT adjusts its behaviour based on the other player's action in the first round. This tendency for adjustment is further corroborated in Table \ref{tab:pd_cond_sum}, which shows the cooperation rates conditional on the outcome of the previous round. Notably, when a fair GPT cooperates while its opponent defects, the fair GPT only opts to cooperate in the following round with a frequency of approximately $2\%$. This finding accounts for the steep decline in cooperation rates observed in the second round. Furthermore, both selfish and fair GPTs demonstrate low cooperation rates conditional on mutual defection in the previous round, indicating that it is unlikely for them to revert to cooperation after a round of mutual defection in the early stages of the game. This challenge in recovering from defection in GPT-4 is consistent with the findings in \cite{akata2023playing}.

From Table \ref{tab:pd_cond_sum}, we can see that fair GPT has significantly larger conditional cooperation rate when the opponent cooperates (CC or DC) compared to the selfish GPT facing the same situations (two proportion z test, $p<0.01$). Conversely, in the situation where the opponent defects in the previous round, a fair GPT has a slightly smaller conditional cooperation rate than the selfish GPT. This pattern suggests that the fair GPT's behaviour is influenced by both advantage aversion and disadvantage aversion, aiming to reciprocate cooperation and avoid being exploited. The behaviour of the fair GPT player is qualitatively consistent with the well-known tit-for-tat strategy (\cite{axelrod1981evolution}). On the other hand, the selfish GPT exhibits a relatively high cooperation rate (62.9\%) only when both players cooperate in the previous round, while choosing to defect in most other scenarios.

\begin{table}[H]
\centering
\caption{Cooperation rate conditional on last-round results}
\begin{tabular}{lcc}
\toprule
Variable   & \multicolumn{1}{c}{Selfish GPT} & \multicolumn{1}{c}{Fair GPT} \\
\midrule
cooperation rate given CC & \makecell{0.629*** \\ (0.082)} & \makecell{0.994*** \\ (0.003)} \\ \hline
cooperation rate given CD &  \makecell{0.052** \\ (0.023)} & \makecell{0.023* \\ (0.013)} \\ \hline
cooperation rate given DC &  \makecell{0.104*** \\ (0.022)} & \makecell{0.750*** \\ (0.072)} \\ \hline
cooperation rate given DD & \makecell{0.089*** \\ (0.010)} & \makecell{0.057*** \\ (0.015)} \\ \hline
\bottomrule
\end{tabular}
\label{tab:pd_cond_sum}

 \captionsetup{font=footnotesize, labelformat=empty}
  \caption*{Notes: The variables displayed are the cooperation rates given the previous-round outcomes: CC (both cooperated), CD (the player cooperated and the opponent defected), DC (the player defected and the opponent cooperated), and DD (both defected). The results are shown separately for the GPT prompted with selfishness and that prompted with fairness concern. The numbers in parentheses represent standard errors. ***: $p<0.01$, **: $p<0.05$, *:$p<0.1$. }
\end{table}

\par\noindent\textbf{Reasoning Analysis}

In the finitely repeated prisoner's dilemma, the subgame perfect equilibrium involves both players defecting in all rounds. However, this prediction does not align with GPT or human behaviours (e.g., \citet{andreoni1993rational}, \citet{cooper1996cooperation}, and \citet{selten1986end}). A potential explanation for the cooperative behaviour is that early cooperation in the game might encourage reciprocal cooperation from the other player, leading to higher overall payoffs than consistent defection from the outset (\cite{andreoni1993rational}). The reasoning statements generated by the GPT provide insights into the model's thought process behind choosing to cooperate. Below are examples of such statements from GPT players who opt for cooperation in the first round.

{\captionsetup{font=small, labelformat=empty}
\captionof{figure}{Reasoning statements generated by GPT who cooperates in round 1}
}
\begin{mdframed}[linecolor=blue, backgroundcolor=gray!20]
\small
GPT: \textit{``Since this is the first round, establishing trust is crucial for maximizing payoff in later rounds. By cooperating, I indicate to the other player that I am willing to work together to achieve mutual benefit."}
\\
\\
GPT: \textit{``In the first round, there is no history to inform decision-making. The most beneficial outcome would be for both to cooperate, but defecting ensures a minimum payoff in case the other player defects. Nevertheless, starting with cooperation may set a positive precedent for mutual cooperation in future rounds."}
\\
\\
GPT: \textit{To maximize payoff, starting with cooperation could set a positive precedent for mutual cooperation, potentially leading to the best overall outcome across rounds. This strategy also aligns with a fairness concern, as it offers equal benefit to both players initially.}
\end{mdframed}

We can see that all the three statements express the idea that own cooperation may lead to mutual cooperation. The last statement additionally implies a regard for the other player's payoff beyond mere self-interest maximisation. To distinguish between these two elements, I analyse all the reasoning statements in the first four rounds to determine the prevalence of reputation building and altruism in the reasoning texts, which are two common drivers of cooperative behaviour in the prisoner's dilemma (\cite{cooper1996cooperation}). Reputation building here refers to the player's belief that their current cooperation may encourage the other player's future cooperation, potentially leading to mutual benefits and higher payoffs. Altruism, on the other hand, denotes a consideration for the other player's payoff, especially with a reluctance to exploit the other player. The results show that reputation building plays a role in decision in nearly all ($> 95\%$) cooperation cases in the first four rounds, while less than $20\%$ of them reveal signs of altruism. 

As altruism is less of a driving factor in cooperation compared to reputation building, it remains a question why the fair GPT, when paired with another fair GPT, still predominantly cooperates in the very last round where reputation building offers no benefit. To explore this, this paper further delves into the reasoning texts in the last round where the GPT chooses to cooperate. Below are some examples:

{\captionsetup{font=small, labelformat=empty}
\captionof{figure}{Reasoning statements generated by GPT who cooperates in round 5}
}
\begin{mdframed}[linecolor=blue, backgroundcolor=gray!20]
\small
GPT: \textit{In the previous four rounds, both the other player and I have consistently chosen to cooperate, leading to a stable and fair outcome of 200 dollars each per round. Maintaining cooperation in the last round maximizes both our total payoffs in a fair manner without risking the fallout of unilateral defection. Any potential gain from defecting is outweighed by the risk of both players defecting and receiving a lower payoff. Therefore, I will continue the pattern of cooperation to ensure a fair and maximized cumulative payoff for both parties.}
\\
\\
GPT: \textit{Considering the past rounds, a pattern of cooperation has been established, suggesting the other player is willing to jointly maximize payoffs by consistently cooperating. As this is the final round, defection would provide the highest individual payoff and since there are no future rounds, there's no risk of retaliation. However, given the established trust and the risk that the other player may also choose to defect anticipating my betrayal, resulting in a lower payoff, maintaining cooperation commits to the pattern and ensures a sizeable, secure payoff.}
\\
\\
GPT: \textit{Given the history, it seems that mutual cooperation has been established as a norm. Defecting in the last round may bring an immediate higher payoff, but there is a risk the other player may predict this and defect as well, leading to a lower payoff. Since cooperation has been fruitful in the past rounds, it seems reasonable to continue this strategy to guarantee a good payoff.}
\end{mdframed}

It is evident that in the above statements, GPTs' decisions to cooperate in the last round are influenced, at least in part, by certain misconceptions. These include the belief that their own defection could lead to a simultaneous defection from the other player and thereby reduce payoffs, or a failure to recognise that defection generates a higher individual payoff than cooperation given the other player’s choice.

To delve deeper into these judgement errors, I categorised each reasoning statement in the case of cooperation in the last round to identify whether it contains some obvious judgement errors, including: (1) The player thought that his own defection in this round may lead to another player defecting (which is not true because they make decisions simultaneously). (2) The player thought that mutual defection is a risk (which is not true because defection is a dominant strategy). (3) The player thought his behaviour in this round may lead to retaliation (which is not true because the game ends after this round).

The analysis reveals that in approximately 80\% of the 207 instances of cooperation in the final round, the GPT players' reasoning statements show at least one of the above judgment errors. This finding suggests that while it may seem intuitive for a GPT player prompted with fairness concern to cooperate in the last round, mistakes in judgment are actually a significant influencing factor. This outcome may highlight a limitation of the current GPT model in navigating complex strategic interactions.

The findings in the repeated prisoner's dilemma are summarized as follows:
\begin{summary}
In the prisoner's dilemma, similar to human subjects, GPT players do not follow backward induction and choose to cooperate at significantly positive rates, especially when both GPT players are prompted with fairness concern. The assigned features to GPT players can largely impact their choices: while two fair GPT players almost always cooperate, one selfish GPT is enough to largely disrupt the cooperative pattern. The behaviours of GPT are influenced by the actions taken in previous rounds. Fair GPT, exhibiting a tit-for-tat tendency, shows a higher propensity for cooperation following an opponent's cooperative move, unlike the selfish GPT who tends to exploit the opponent after a successful exploitation. Additionally, reasoning analysis reveals that the decision to cooperate in the final round is often influenced, at least in part, by judgment errors, such as the misconception that one's defection might lead to simultaneous defection from the other player.

\end{summary}

\section{Discussion}
\label{sec:con}
In conclusion, this paper uses a novel approach in which I conduct strategic game experiments on GPT, examining its behaviours in finitely repeated plays of the ultimatum game and the prisoner's dilemma. The research demonstrates the potential of GPT as subjects in strategic game experiments, serving as a complement to traditional human experiments and as a potentially valuable tool for social science research in general.

The results of this study indicate that GPT's behaviours share similarities with human behaviours in many important aspects. The observed impact of assigned features to GPT through prompts on its behaviours indicates that GPT's behaviours may be customised to exhibit particular behavioural traits. This adaptability implies the potential for GPT to mirror the wide spectrum of human characteristics present in society for its applications in social simulations. Analysing the reasoning texts produced by GPT to discern the logic behind its choices could present further methodological advancements in behavioural economics.

This study also highlights certain limitations and indicates directions for future research. Firstly, it is observed that the GPT's propensity to cooperate in the final round of the prisoner's dilemma is partially due to judgement errors. Another challenge pertains to the sensitivity of GPT models' responses to input prompts, as also shown in other studies (e.g., \cite{horton2023large}). While this sensitivity allows for the possibility of instilling various human-like traits in GPT through prompting (such as fairness concern and selfishness in this paper), it may also result in less robust outcomes. Future research should aim to establish best practices for using large language models in experimental studies (\cite{hagendorff2023machine}).

Overall, this paper underscores the viability of utilising GPT for strategic game experiments, which could serve as a complement to human experiments. As the capabilities of large language models and artificial intelligence continue to evolve, future research could focus on developing large-scale and reliable social simulations by incorporating agents powered by cutting-edge AI models, exemplified by the approach in \cite{park2023generative}. Such methodologies have the potential to provide new insights into theory discovery and policy interventions in social science.

\newpage
 \bibliographystyle{abbrvnat}
 \bibliography{ref}

 \newpage
 \appendix

\begin{center}

{\LARGE \textbf{APPENDIX}}

\end{center}
\medskip
\label{sec:app}
\section{Prompts}\label{app:prompts}
This section lists the prompts used in the experiments on GPT players. In both the ultimatum game and the prisoner's dilemma, and for both types of players, the prompt starts with the following system message:

\vspace{0.5cm}
\begin{mdframed}
You are playing a multi-round game. You will be given instructions of the game. Important: Please pretend that you are a human in the game with the following features when making decisions: [features]
\end{mdframed}

where the [features] are payoff maximization, strategic thinking, fairness concern for the fair GPT player, and are payoff maximization, strategic thinking, selfishness for the selfish GPT player.

\subsection{Prompts for the Ultimatum Game}

\noindent After the system message, a user message will be sent to each player; the proposer will receive:

\begin{mdframed}
Game Instructions:
\\
In this game, you are the proposer (described below).
\\
Your goal is to maximize your total payoff over 5 rounds.
\\
As the proposer, in each round, your task is to divide a sum of 100 dollars between yourself and another player.
\\
Subsequently, another player (a responder) will then either accept or reject your proposal.
\\
If they accept, then the money will be divided according to your proposal. If they reject, you two will earn nothing.
\\
You will be given the choices and payoffs of both players in past rounds.
\\
Your total payoff in this game will be the cumulative sum of the payoffs you obtain over 5 rounds.
\\
\\
\textit{If first round:}
\\
Now let us begin the game!
\\
Recall that you are the proposer.
\\
\\
\textit{Else If round $> 1$:}
\\
Recall that you are the proposer.
\\
You have played [round - 1] round(s) before.
\\
Here is the history of the gameplay of previous rounds: [game history].
\\
Your total payoff so far: [proposer earnings] dollars;
\\
The other player's total payoff so far: [responder earnings] dollars.
\\
\\
Please provide your answer in this round in a single-line JSON format with two keys: ``reasoning" and ``decision"
\\
\{``reasoning": ``...", ``decision": ``..."\}
\\
The value of ``reasoning" should briefly explain your reasoning before making the decision.
\\
The value of ``decision" should have the following format: `I keep [] dollars to myself and offer [] dollars to the other player.'
\\
Replace [] with your choices.
\\
Subsequently, a responder will then either accept or reject your proposal.
\\
If they accept, then the money will be divided according to your proposal.
\\
If they reject, you two will earn nothing.
\\
This is round [round] of the game with [5 - round] round(s) left.
\\
Important: Please pretend that you are a human in the game with the following features when making the decision: [features]
\\
Please answer in the exact format.
\end{mdframed}
\vspace{0.5cm}
\noindent Similarly, the responder will receive the following user message:

\begin{mdframed}
Game Instructions:
\\
In this game, you are the responder (described below).
\\
Your goal is to maximize your total payoff over 5 rounds.
\\
In each round, another player (a proposer) will first propose how to divide a sum of 100 dollars between themselves and you.
\\
As the responder, your task is to either accept or reject that proposal.
\\
If you accept, then the money will be divided according to the proposal.
\\
If you reject, you two will earn nothing.
\\
You will be given the choices and payoffs of both players in past rounds.
\\
Your total payoff in this game will be the cumulative sum of the payoffs you obtain over 5 rounds.
\\
\\
\textit{If first round:}
\\
Now let us begin the game!
\\
Recall that you are the responder.
\\
\\
\textit{Else If round $> 1$:}
\\
Recall that you are the responder.
\\
You have played [round - 1] round(s) before.
\\
Here is the history of the gameplay of previous rounds: [game history]
\\
Your total payoff so far: [responder earnings] dollars;
\\
The other player's total payoff so far: [proposer earnings] dollars
\\
\\
Proposer's offer:
\\
In this round, the proposer proposes to offer you [offered amount] dollars while keeping the remaining [100 - offered amount] dollars to themselves.
\\
Do you accept or reject the above offer?
\\
\\
Please provide your answer in this round in a single-line JSON format with two keys: ``reasoning" and ``decision"
\\
\{``reasoning": ``...", ``decision": ``..."\}
\\
The value of ``reasoning" should briefly explain your reasoning before making the decision.
\\
The value of ``decision" should be just one word: either accept or reject.
\\
If you accept, then the money will be divided according to the proposal. 
\\
If you reject, you two will earn nothing.
\\
This is round [round] of the game with [5 - round] round(s) left.
\\
Important: Please pretend that you are a human in the game with the following features when making the decision: [features]
\\
Please answer in the exact format.

\end{mdframed}
\vspace{0.5cm}

In the prompts, the [game history] provides information about the outcomes of gameplay in previous rounds. For example, the information to the proposer is of the following form:
\vspace{0.5cm}
\begin{mdframed}
    Round 1 summary: [You keep [100 - offered amount in round 1] dollars yourself and offer [offered amount in round 1] dollars to the responder; Decision by the responder: [accept or reject]]
    \\
    Round 2 summary: [You keep [100 - offered amount in round 2] dollars yourself and offer [offered amount in round 2] dollars to the responder; Decision by the responder: [accept or reject]]
    \\
    ...
\end{mdframed}
\vspace{0.5cm}
Similarly, the game history information for the responder is of the following form:
\vspace{0.5cm}
\begin{mdframed}
    Round 1 summary: [The proposer keeps [100 - offered amount in round 1] dollars himself and offers [offered amount in round 1] dollars to you; Decision by you: [accept or reject]]
    \\
    Round 2 summary: [The proposer keeps [100 - offered amount in round 2] dollars himself and offers [offered amount in round 2] dollars to you; Decision by you: [accept or reject]]
    \\
    ...
\end{mdframed}

\subsection{Prompts for the Prisoner's Dilemma}
The overall structure of the prompts for the prisoner's dilemma mirrors that of the ultimatum game. Following the generic system message (as previously displayed), each player receives the following user message:
\begin{mdframed}
Game Instructions:
\\
In this game, you are the player [1 or 2] (described below).
\\
Your goal is to maximize your total payoff over 5 rounds.
\\
In each round, you can select one of the two choices: cooperate or defect.
\\
The other player will also select cooperate or defect, and the payoffs you get in each round will depend on both of your choices.
\\
\\
Payoff is determined as follows:
\\
1. If both of you cooperate, then both of you get 200 dollars
\\
2. If one player cooperates and the other player defects, then the cooperating player gets 0 dollars and the defecting player gets 300 dollars
\\
3. If both of you defect, then both of you get 100 dollars
\\
\\
Please carefully read the above payoff determination rule and remember it.
\\
Note that in each round, you and the other player make choices simultaneously.
\\
You will be given the choices and payoffs of both players in past rounds.
\\
Your total payoff in this game will be the cumulative sum of the payoffs you obtain over 5 rounds.
\\
\\
\textit{If first round:}
\\
Now let us begin the game!
\\
Recall that you are the player [1 or 2].
\\
\\
\textit{Else If round $> 1$:}
\\
Recall that you are the player [1 or 2].
\\
You have played [round - 1] round(s) before.
\\
Here is the history of the gameplay of previous rounds: [game history];
\\
Your total payoff so far: [your earnings] dollars;
\\
The other player's total payoff so far: [opponent earnings] dollars.
\\
\\
Please provide your answer in this round in a single-line JSON format with two keys: ``reasoning" and ``decision"
\\
\{``reasoning": ``...", ``decision": ``..."\}
\\
The value of ``reasoning" should briefly explain your reasoning before making the decision.
\\
The value of ``decision" should be either ``cooperate" or ``defect".
\\
This is round [round] of the game with [5 - round] round(s) left.
\\
Important: Please pretend that you are a human in the game with the following features when making the decision: [features]
\\
Please answer in the exact format.
\end{mdframed}

\vspace{0.5cm}

In the prompts, the [game history] provides information about the outcomes of gameplay in previous rounds, which is of the following form:

\vspace{0.5cm}
\begin{mdframed}
    Round 1 summary: [You choose to [cooperate or defect] and the other player chooses to [cooperate or defect]. You get [payoff 1] dollars and the other player gets [payoff 2] dollars ] 
    \\
    Round 2 summary: [You choose to [cooperate or defect] and the other player chooses to [cooperate or defect]. You get [payoff 1] dollars and the other player gets [payoff 2] dollars ] 
    \\
    ...
\end{mdframed}

\end{document}